\documentclass[doublecol]{epl2} 

\usepackage{amssymb}

\title{Fast Quantum Communication in Linear Networks}
\author{Kurt Jacobs\inst{1,2,3} \and Rebing Wu\inst{4} \and Xiaoting Wang\inst{3} \and Sahel Ashhab\inst{5} \and Qi-Ming Chen\inst{4} \and Herschel Rabitz\inst{6}}
\shortauthor{K. Jacobs \etal}

\institute{                    
  \inst{1} U.S. Army Research Laboratory, Computational and Information Sciences Directorate, Adelphi, Maryland 20783, USA\\
  \inst{2} Department of Physics, University of Massachusetts at Boston, Boston, MA 02125, USA \\
  \inst{3} Hearne Institute for Theoretical Physics, Louisiana State University, Baton Rouge, Louisiana 70803, USA\\
  \inst{4} Department of Automation, Tsinghua University \& Center for Quantum Information Science and Technology, TNList, Beijing, 100084, China \\
  \inst{5} Qatar Environment and Energy Research Institute (QEERI), HBKU, Qatar Foundation, Doha, Qatar \\
  \inst{6} Department of Chemistry, Princeton University, Princeton, NJ 08544, USA 
}


\abstract{
Here we consider the speed at which quantum information can be transferred between the nodes of a linear network. Because such nodes are linear oscillators, this speed is also important in the cooling and state preparation of mechanical oscillators, as well as frequency conversion. We show that if there is no restriction on the size of the linear coupling between two oscillators, then there exist control protocols that will swap their respective states with high fidelity within a time much less than a single oscillation period. Standard gradient search methods fail to find these fast protocols. We were able to do so by augmenting standard search methods with a path-tracing technique, demonstrating that this technique has remarkable power to solve time-optimal control problems, as well as confirming the highly challenging nature of these problems. As a further demonstration of the power of path-tracing, first introduced by Moore-Tibbets~\textit{et al.} [Phys. Rev. A 86, 062309 (2012)], we apply it to the generation of entanglement in a linear network.  
} 

\begin{document}

\newcommand{\nn}{\nonumber} 
\newcommand{\mbs}[1]{\mbox{\scriptsize #1}}
\newcommand{\mbsi}[1]{\mbox{\scriptsize\textit{#1}}}
\newcommand{\dg}{^\dagger}
\newcommand{\smallfrac}[2]{\mbox{$\frac{#1}{#2}$}}
\newcommand{\ket}[1]{| {#1} \ra}
\newcommand{\bra}[1]{\la {#1} |}
\newcommand{\pfpx}[2]{\frac{\partial #1}{\partial #2}}
\newcommand{\dfdx}[2]{\frac{d #1}{d #2}}
\newcommand{\half}{\smallfrac{1}{2}}

\maketitle 

\section{Introduction}

The transfer of quantum information between harmonic oscillators coupled by linear interactions has a variety of applications, including communication in linear networks~\cite{Chudzicki10}, cooling of mechanical resonators~\cite{Wang11, Machnes12}, and frequency conversion~\cite{Tian10, Stannigel10, Regal11, Safavi11b, Tian12, Wang12}. For cooling mechanical oscillators, transferring the entropy of the mechanical oscillator to a superconducting or optical oscillator is the most effective method known to-date~\cite{Wilson-Rae07, Marquardt07, Tian09, Schliesser08, Teufel11b}. The faster the transfer can be performed the lower the achievable temperature, or equivalently the higher the resulting purity of the prepared state~\cite{Wang11, Machnes12}. Frequency conversion is achieved by transferring the state of an oscillator at one frequency to an oscillator at a different frequency. While here we focus on communication in linear networks, our results are applicable to all the above applications. Speed is important for obvious reasons in communication and computation, and it is of additional importance in quantum technologies because of the ever-present effects of environmental noise that degrade quantum states over time. Here we are concerned with finding time-dependent control protocols that will transfer information at the maximum speed, a problem more generally referred to as ``time-optimal'' control~\cite{Anandan90, Giovannetti03, Margolus98, delCampo13, Carlini06, Carlini07, Hegerfeldt13}. 

Two oscillators A and B that are coupled by a linear interaction are described by the Hamiltonian 
\begin{equation} 
 H = H_0 + V_{\mbs{I}}, 
\end{equation} 
with 
\begin{eqnarray} 
 H_0 & = & \hbar \omega_a a^\dagger a + \hbar \omega_b b^\dagger b ,  \\
 V_{\mbs{I}} & = & \hbar g x_a x_b , 
\end{eqnarray} 
where $x_a = a + a^\dagger$,  $x_b = b + b^\dagger$, $w_a$ and $w_b$ are the respective frequencies of A and B, and $g$ is the linear coupling rate~\footnote{There is no need to choose the more general form of the linear interaction in which $x_a = e^{- i \theta} a + e^{i \theta}a^\dagger$, $x_b = e^{- i \phi} b + e^{i \phi}b^\dagger$, since varying $\theta$ and $\phi$ merely shifts the relative phases of the oscillators.}. Given this interaction it is not obvious how to engineer a unitary transformation that will transfer an arbitrary state of A to B. Note that for the purposes of information transfer, an operation transfers an arbitrary state $|\psi\rangle$ of A to B if for every $|\psi\rangle$ the oscillator B finishes in the state $U |\psi\rangle$ where $U$ is unitary that does not depend on $|\psi\rangle$. If A and B form a closed system, then this unitary must swap their respective states (up to the local evolution of either oscillator) because unitary operations are reversible; B's state has to go somewhere when replaced by A's.  

A swap between A and B can be obtained in an especially simple way if the coupling rate $g$ is much smaller than the frequencies $\omega_a$ and $\omega_b$. To do this one first modulates the coupling rate at the difference frequency $\Delta = |\omega_a - \omega_b|$ so that the interaction Hamiltonian becomes $V_I = g\cos(\Delta t) x_a x_b$. Moving into the interaction picture with respect to $H_0$ and making the rotating-wave approximation the Hamiltonian becomes \begin{equation}
  H_{\mbs{I}} = \hbar g (b^\dagger a +  a^\dagger b) , 
\end{equation}
where the subscript denotes the interaction picture. Remarkably this interaction swaps the states of the resonators in a time $\tau_{\mbs{RWA}} = \pi/g$. The speed of this method of state-swapping is restricted, however, by the requirement that $g \ll \mbox{min}(\omega_a,\omega_b)$, and thus $\tau_{\mbs{RWA}} \gg T_{\mbs{slow}}$ where $T_{\mbs{slow}}$ is the period of the slower of the two oscillators. 

In 2011 Wang \textit{et al.}~\cite{Wang11} and Machnes \textit{et al.}~\cite{Machnes12} showed that a swap between two linear oscillators could be achieved within a single oscillation of the slower oscillator by using a time-dependent modulation of the coupling constant $g$. The technique of engineering unitary operations by changing the Hamiltonian with time is an important one within the toolset of \textit{quantum control}~\cite{Brif10, Schulte05}, and a given prescription for varying the Hamiltonian is called a \textit{control protocol}. While this technique is powerful, finding control protocols to perform a given task is a highly complex problem, and one for which numerical search methods are often essential~\cite{Kosloff89, Schulte05, Grace07, Haidong09, Doria11, Nimbalkar12, Ashhab12, Huang14}. 

In the case of two coupled oscillators the unitary operations that can be realized by varying the coupling rate $g$ are those within the algebra generated by the three operators $a^\dagger a$, $b^\dagger b$, and $x_a x_b$, as these are the three operators that are effectively combined with different weights by the variation. Since the Hamiltonian of the slower oscillator is essential for generating a swap, we can expect that its frequency will limit the swapping rate --- there must be enough time for this Hamiltonian to make a non-trivial contribution to the dynamics. By using a numerical search Wang \textit{et al.}~\cite{Wang11} found a control protocol that performed a swap in a little over half the period $T_{\mbs{slow}}$. By starting the numerical search with an approximate analytical protocol Machnes \textit{et al.}~\cite{Machnes12} were able to obtain a swap in a little over a quarter of $T_{\mbs{slow}}$. It appeared unlikely from these results that significantly shorter swap times were possible. 

Here we show that, while the frequency of the slow oscillator does ultimately limit the speed of a swap operation, high-fidelity swaps can be performed in times significantly shorter that those previously known. Finding these fast protocols was made possible by a ``path-tracing'' technique that we describe below. While path tracing itself is not new --- it was introduced by Moore-Tibbets~\textit{et al.}~\cite{Moore12b} as a fast method to find protocols when one wishes to scan across protocol durations --- its ability to readily solve problems that are virtually impossible otherwise has not been previously demonstrated. An optimization problem becomes hard when local minima and/or saddle points become so dense that gradient search methods get trapped at these points with high probability. In finding previously undiscovered state-swapping protocols for oscillators we demonstrate two key properties of quantum control: i) quantum control problems that are easy for long protocol times~\cite{Wu08} can become very hard as the protocol duration is reduced, with the result that time-optimal quantum control problems can be similarly hard; ii) path tracing is able to solve at least some of these hard problems. 

\section{The Method of Path Tracing}
\label{pathtrace}

Before we describe path tracing we introduce some terms and definitions. A numerical search method is a procedure for finding a local minimum of a function. If the local minimum found is also a global minimum then the search has solved the minimization problem. To find a control protocol that achieves a specific unitary transformation $V$ we use a search method to minimize a quantity, $\varepsilon$, that measures the difference between $V$ and the unitary generated by the protocol. For example, if the protocol generates the unitary $U(T)$ over a time $T$, then 
\begin{equation}
   \varepsilon = 1 - \mbox{Tr}[V^\dagger U(T)] 
\end{equation}
is a good measure of the difference between $U(T)$ and $V$. Because the search method must minimize $\epsilon$ it cannot simultaneously minimize the time that the protocol takes, and this is an essential difficulty in time-optimal control. Let us define a ``good protocol'' as one with an error $\epsilon$ that is below some threshold $\varepsilon_{\mbs{t}} \ll 1$. The standard approach to finding a protocol with minimum time is to perform a number of independent searches, where each search uses a different fixed time $T$. In this way one obtains protocols for a range of durations. It is then hoped that of the good protocols so obtained, the one with the smallest value of $T$ is near-optimal. For problems for which this procedure works well, one finds that above a critical time $T_{\mbs{min}}$ the resulting error $\varepsilon$ is zero (to within machine precision), and below this time the error steadily increases away from zero. The fact that the behavior of $\varepsilon$ is a continuous function of $T$ provides confidence that $T_{\mbs{min}}$ is the minimum time for which perfect protocols can be obtained. For the problems considered here we do not find this behavior; instead there is a critical time $T_{\mbs{cr}}$ at which the error obtained by the search jumps abruptly from zero to a large value. This behavior is a clear sign that the search is being trapped in local minima below $T_{\mbs{cr}}$.

Even when the above procedure works well, it is certainly cumbersome. To increase the efficiency of the process Moore-Tibbets~\textit{et al.}~\cite{Moore12b} introduced the following procedure. i) Start with a duration $T_0$ that is expected to be larger than the minimum time $T_{\mbs{min}}$ and obtain a protocol $p(T_0)$. ii) Reduce the duration to $T_1 = T_0 - \Delta T$ and perform a new search, but this time starting the search at the protocol $p(T_0)$ found in i), where this protocol is now allowed to run only for the duration $T_1$. In particular, we simply scale the protocol $p(T_0)$ along the time axis so that it takes time $T_1$, producing a slightly modified unitary transformation. iii) Repeat this procedure, on each iteration reducing the time by $\Delta T$, and starting the search for this new duration with the protocol found in the previous step. In this way, each time we search for a new protocol we can expect to be starting the search at a protocol that is close to the one we seek, thus reducing the search time. We refer to this procedure as ``path tracing'' because we expect the sequence of protocols obtained to trace a continuous path in protocol space. Moore-Tibbets~\textit{et al.} applied this method to problems for which the standard method works well. They found that their path-tracing method was indeed significantly faster than the standard method, successfully tracing a path of minimum error as a function of duration, $\varepsilon_{\mbs{min}}(T)$. The curve given by $\varepsilon_{\mbs{min}}(T)$ is called the fidelity-time tradeoff frontier. 

We write the Hamiltonian for a control problem in the form 
\begin{equation}
   H(t) = H_0 + \sum_{j=1}^M \lambda_j(t) H_j ,
\end{equation}
in which $H_0$ and $\{H_j\}$ are fixed and $\{\lambda_j(t)\}$ is the set of parameters that are available to be varied as functions of time. These functions are referred to as the control functions, and the rate at which they must be changed with time in order to implement a given control protocol is an important practical consideration. Because of this, in searching for good control protocols we wish to bound this rate. We must also discretize the control functions so that we have a finite set of parameters over which to minimize $\varepsilon$. If we discretize each of the control functions using $N$ parameters, then we may write 
\begin{equation}
   \lambda_j(t) = \sum_k \lambda_{jk}f_{jk}(t), \;\;\;\; k = 1,\ldots, N ,  
\end{equation}
in which $f_{jk}(t)$ are fixed functions of time, and $\{ \lambda_{jk} \}$ is the set of $MN$ parameters over which to optimize. A natural way to discretize the control functions when limiting their rate of change (bandwidth) is to represent each by a truncated Fourier series. Here we instead use a simpler discretization which is effectively equivalent. We make the control functions piecewise-constant on $N$ time segments each of length $T/N$. This choice is ideal for explorations because it allows us to solve for the evolution using matrix exponentiation, which provides a robust solution without any time-stepping error, and is relatively fast for  dimensions that are not too large. While the discontinuities in the control functions mean that technically they have infinite bandwidth, experience shows that the existence of a piecewise-constant protocol implies the existence of one or more continuous protocols that have the same error and the same number of degrees of freedom $\lambda_{jk}$ (see, e.g.~\cite{Wang11}). If we represent the control functions using a Fourier series, then the number of degrees of freedom is $MN$, where in this case $N$ is the number of terms in the Fourier series. The minimum bandwidth is then approximately $N/(2T)$. Finding a piecewise-constant protocol with $N$ segments thus implies that there exists a similar protocol with a bandwidth $\sim N/(2T)$. In performing the path-tracing procedure we will keep the number of intervals, $N$, fixed, thus decreasing the length of each interval as we reduce $T$. 

\begin{figure}[t]
\begin{center}
\onefigure[width=8.2cm]{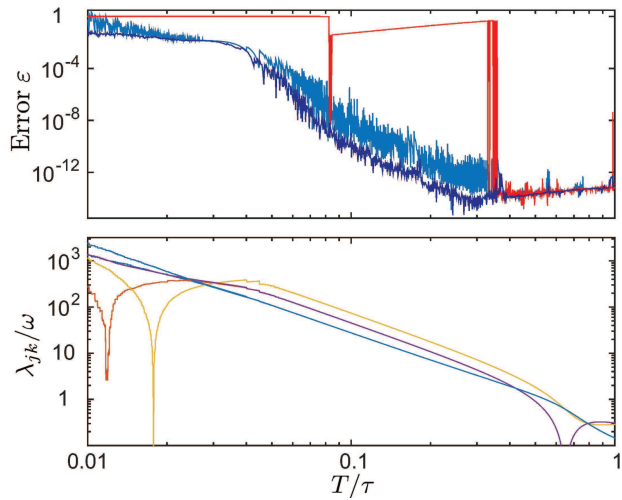}
\vspace{-0.9cm}
\end{center}
\caption{(Color online) Protocols that swap the states of two linearly-coupled oscillators. The slower oscillator has period $\tau = 2\pi/\omega$. (a) The error $\varepsilon$ as a function of duration, $T$. Red: protocols found using independent gradient searches for each of 2000 values of the duration, $T$. Light blue: protocols found using a gradient search to trace a single path through decreasing time, using 2000 points on the path. Dark blue: protocols with the least error of those found by tracing 11 paths. A total of 64845 points were used in tracing these 11 paths, for an average of 5895 points per path. (b) The five control parameters as a function $T$ for the protocols whose errors are shown in the light blue curve in (a).} 
\label{fig1} 
\end{figure} 

\section{State-transfer in linear networks}
\label{apps}

We now apply the path-tracing method to the transfer of a quantum state from one harmonic oscillator to another. We consider two configurations: in the first the oscillators A and B are directly coupled, and in the second each is instead coupled to a third oscillator, C. The numerical simulation of these scenarios is greatly facilitated by the fact that Gaussian states remain Gaussian under linear evolution. Because of this, if the mean values of the canonical coordinates are zero, we need only keep track of their second moments. For two linearly coupled oscillators with respective annihilation operators $a$ and $b$, the equation of the motion for joint second moments is given by \begin{equation}
   \frac{dC}{dt} = A C + C A^{\mbs{t}} , 
\end{equation}
where 
\begin{equation}
   C \equiv \langle \mathbf{v}^{\mbox{\scriptsize t}}\mathbf{v}\rangle \;\;\;\; \mbox{with} \;\;\;\; \mathbf{v} \equiv (a,a^\dagger,b,b^\dagger) , 
\end{equation}
and 
\begin{equation}
   A =  \left( \begin{array}{cccc}  \omega & 0 &  -i\lambda_1(t) g &  -i\lambda_1(t) g  \\  0 &  -\omega  & i\lambda_1(t) g &  i \lambda_1(t) g  \\ -i\lambda_1(t) g & -i\lambda_1(t) g & \omega & 0 \\ i\lambda_1(t)  g & i\lambda_1(t) g & 0 & -\omega  \end{array} \right) , 
\end{equation}
in which the single control parameter is $\lambda_1$. Since the evolution is unitary, a transfer of von Neumann entropy $S$ is equivalent to the transfer of $S$ qubits of quantum information~\cite{MikeandIke}. To simulate this transfer we therefore start oscillator A in a mixed Gaussian state (we choose the state with $\langle a \rangle = \langle a^\dagger \rangle = 0$ and $\langle a^\dagger a \rangle = 1$, which has von Neumann entropy $S = 2~\mbox{bits}$) and oscillator $B$ in the ground state. The condition for successful transfer is that oscillator A is left in the ground state so that all the entropy is stored in B. Since the oscillators are undriven the ground state is the only accessible pure state, thus in achieving the swap the control protocol is free to apply any local unitary to either oscillator. 

For the numerical search we define the error as $\varepsilon = \langle a^\dagger a \rangle$, divide the duration $T$ into $N=5$ segments (giving five parameters $\lambda_{1k}$, $k=1,\ldots,5$), and use the BFGS gradient search method~\cite{Nocedal06}. The results are shown in Fig.~\ref{fig1}. We see that  the standard search method finds no good protocols below $T \approx \tau/3$ where $\tau \equiv 2\pi/\omega$ (with a brief exception at $T \approx 0.08 \tau$). When we instead trace a single path a very broad trade-off frontier is revealed in which the achievable error increases slowly as $T$ is reduced. We also show the trade-off frontier that results from taking the best protocols over 11 paths. That the resulting frontier is smooth and that we find no paths that are significant outliers are indications that it may be the true frontier for this control task. If we set our error tolerance at $\varepsilon = 10^{-4}$ then we can perform the swap in just under $T = \tau/20$. 

\begin{figure}[t]
\begin{center}
\onefigure[width=8.2cm]{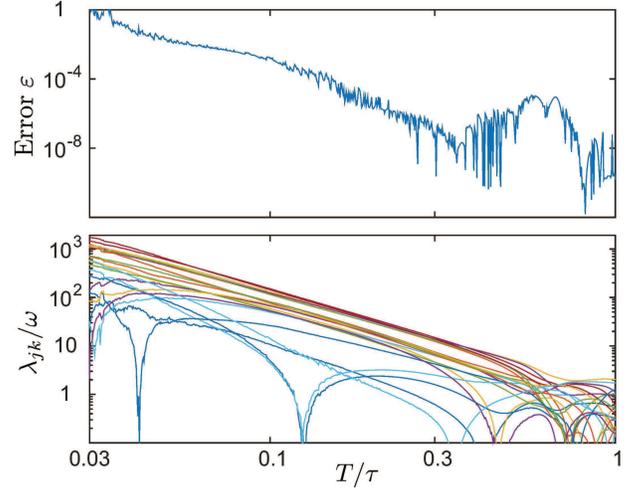}
\vspace{-0.9cm}
\end{center}
\caption{(Color online) Protocols that swap the states of two oscillators that are linearly coupled via a third oscillator. The slower oscillator has period $\tau = 2\pi/\omega$. (a) The error $\varepsilon$ as a function of duration, $T$, for protocols found by tracing a single path through decreasing time, using 900 points on the path. (b) The 20 control parameters as a function of $T$ for the protocols in (a).} 
\label{fig2} 
\end{figure} 

In Fig.~\ref{fig2} we show results for the task of transferring a state from A to B when the two oscillators are linearly coupled via a third oscillator C. Note that we can perform this operation using two state-swapping operations for directly coupled oscillators (swap A with C and then B with C). We are therefore interested in whether it is possible to perform the transfer A to B in less than twice the time of the previous protocol with no more than twice the error. For this scenario we start B and C in the ground state and the ideal final state is that A and C are both in the ground state. This time we allow the controller to vary the coupling rates between both pairs of oscillators, and use 10 time segments for the a total of 20 control parameters. In this case independent searches can take so long that it is impractical to obtain protocols in this way. Path tracing is much faster, and we see from the single path shown in Fig.~\ref{fig2}, that if we allow an error of $\varepsilon = 2\times 10^{-4}$, the time taken by the fastest protocol that achieves this error is approximately $T = 2\tau/20$. This is no faster than two swap protocols performed in sequence. 

\begin{figure}[t]
\begin{center}
\onefigure[width=8.2cm]{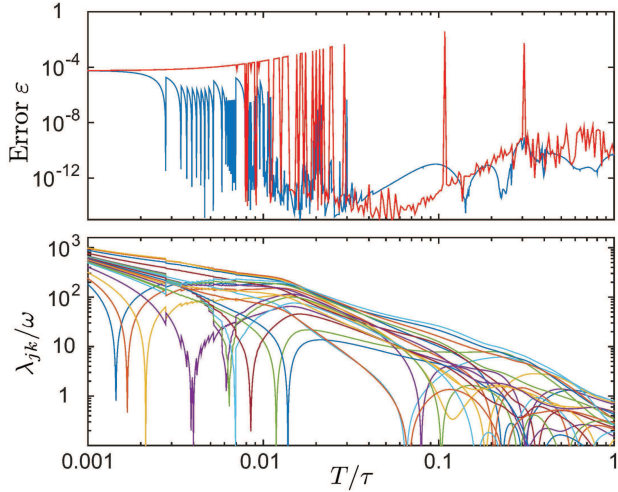}
\vspace{-0.9cm}
\end{center}
\caption{(Color online) Protocols that generate an entangled two-mode squeezed state, given in the main text, between two oscillators linearly coupled via a third oscillator. The slower oscillator has period $\tau = 2\pi/\omega$. (a) The error $\varepsilon$ as a function of duration, $T$. Red: protocols found using independent gradient searches for each of 400 values of $T$. Light blue: protocols found by tracing a single path through decreasing time, using 1350 points on the path. (b) The 20 control parameters as a function $T$ for the protocols whose errors are shown in the light blue curve in (a).} 
\label{fig3}  
\end{figure} 

\section{Generation of entanglement}

As an additional example of the use of path tracing, we consider the generation of entanglement between two oscillators that are again coupled only via their interactions with a third oscillator. This control problem provides an especially clear demonstration of trapping in local minima as $T$ is reduced. For this task all three oscillators start in the ground state and we wish to prepare a two-mode squeezed state between oscillators A and B. This state is defined by 
\begin{equation}
  \langle a^\dagger a\rangle = \langle b^\dagger b\rangle = \frac{\cosh(2r)}{2}, \;\;\;\;\; \langle a b^\dagger \rangle = \frac{\sinh(2r)}{2}, 
\end{equation}
and has an entanglement of 
\begin{equation}
  E = (1+\lambda)\ln(1+\lambda) - \lambda\ln\lambda  
\end{equation}
with $\lambda = \sinh^2r$. We define the error as
\begin{eqnarray}
   \varepsilon & = & \left(\langle a^\dagger a\rangle - \frac{\cosh(2r)}{2}\right)^{\!\! 2} + \left(\langle b^\dagger b\rangle - \frac{\cosh(2r)}{2}\right)^{\!\! 2} \nonumber \\ 
   & & + \left(\langle a b^\dagger \rangle - \frac{\sinh(2r)}{2}\right)^{\!\! 2}  \!\! . 
\end{eqnarray}
We choose $r=2$ corresponding to an entanglement of $5.2$ bits. We plot the results in Fig.~\ref{fig3} from which we see that as $T$ is reduced the independent searches, shown in red, jump between poor solutions and good solutions, where trapping in the poor solutions dominates for $T \lesssim 0.01 \tau$. In comparison, the traced path usually remains closer to the good solutions; for small $T$ it climbs repeatedly out of good solutions, but continually drops back into them, something that the independent searches fail to do. 

\section{Summary}
\label{conc}

Here we have used the method of path-tracing, a simple procedure that can be executed easily with any gradient search method, to find previously unknown control protocols for linear quantum networks. In doing so we have shown that path tracing is extremely powerful, finding fast protocols that appear to be out of reach of previous methods. While we have considered here purely oscillator-based problems in the ultra-strong coupling regime, path tracing may well change what is possible with quantum control across a wide range of problems. Preliminary investigations indicate that problems in which sets of qubits are used to interface with, or control the dynamics of resonators become very hard when maximal speed is sort, and are thus natural candidates for the application of this method. 

\acknowledgments In the early part of this work KJ was partially supported by the NSF Project Nos.\ PHY-1005571 and PHY-1212413. HR was partially supported by the ARO MURI grant W911NF-11-1-0268, and RBW by the NSFC Grant Nos.\ 60904034, 61374091, and 61134008. XW acknowledges support from the NSF Project No.\ CCF-1350397.


\end{document}